\documentclass[withtimes]{easychair}

\usepackage{amsmath,amsfonts}

\begin{document}
\title{Linear Models of Computation and Program Learning}
\titlerunning{Linear Models of Computation}
\author{Michael Bukatin\inst{1}\and Steve Matthews\inst{2}}

\institute{Nokia Corporation\\
Burlington, Massachusetts, USA\\ 
\email{bukatin@cs.brandeis.edu}
\and
Department of Computer Science\\
University of Warwick\\
Coventry, UK\\
\email{Steve.Matthews@warwick.ac.uk}}
\authorrunning{Bukatin and Matthews}

\maketitle

\begin{abstract}
We consider two classes of computations which admit taking linear combinations of execution runs:
probabilistic sampling and generalized animation. We argue that the task of program learning
should be more tractable for these architectures than for conventional deterministic
programs. We look at the recent advances in the ``sampling the samplers" paradigm in
higher-order probabilistic programming. We also discuss connections between
partial inconsistency, non-monotonic inference, and vector semantics.

\end{abstract}

\section{Introduction}

One of the key challenges of program learning is that software tends to be too brittle and insufficiently robust with respect to minor variation.

Biological systems tend to be much more flexible and adaptive with respect to variation. In particular, biological cells are capable of functioning
at wide ranges of the level of expression of various proteins, which are machines working in parallel. Regulation of the level of expression
of specific proteins is a key element of flexibility of biological systems. It is argued in evolutionary developmental biology
that the flexible architecture together with conservation of core mechanisms is crucial for the observed rate of 
biological evolution~\cite{JGerhartKirschner,MKirschnerGerhart}.
It is suggested that morphology evolves largely by altering the expression of functionally conserved proteins~\cite{SCarrol}.

To incorporate regulation of expression into a system of genetic programming one might evolve programs describing systems of parallel computational
processes. Then one might take the CPU allocation and other computational resources given to a particular computational process as
computational equivalent of the level of expression of a particular protein. 

Of course, many of the architectures for parallel computations are brittle as well, with delicate mechanisms of writing to shared memory and
locks. To achieve flexibility one should use parallel architectures which minimize those delicate interdependencies.

Computational architectures which admit the notion of linear combination of execution runs are particularly attractive in this sense. 
Then one can regulate the system simply by controlling coefficients in a linear combination of its components.

In this paper we consider two computational architectures which admit linear combinations of execution runs.

One such architecture is probabilistic sampling. If one has two samplers generating points of two distributions with
a uniform speed, so that the notion of ``a number of points generated per unit of time" is well defined for both of them,
one can obtain a sampler generating a linear combination of those two distributions with arbitrary positive
coefficients simply by running those two samplers in parallel with appropriate relative speeds.

We argue that it should be easier to learn probabilistic programs than to learn
deterministic programs due to the fact that probabilistic programs admit linear combinations of execution runs.
We also discuss the techniques to allow negative coefficients in those linear combinations later in the text.

There is a lot of affinity between methods of evolutionary programming and methods of probabilistic
sampling. Evolutionary schemas can be considered as particular sampling methods, while many sampling
schemas have strong evolutionary flavor (more details in Section~\ref{evolution_vs_sampling}).

This means that instead of thinking in terms of genetic programming for probabilistic programs
one might think about program learning in terms of the ``sampling the samplers" paradigm, namely in terms of probabilistic
programs sampling other probabilistic programs (generative models producing other generative
models as their points). 
This  ``sampling the samplers"  paradigm manifests itself, in particular, 
in recent work on learning probabilistic programs by Yura Perov and Frank Wood~\cite{YPerovWood}
and also in recent advances in compositional concept learning obtained by Brenden Lake~\cite{BLake}.
We review these and some other recent advances in higher-order probablistic programming
in Section~\ref{prob_prog}.

\subsection{Negative Coefficients}

Another computational architecture which admits linear combinations of execution runs is generalized animation.
We define a generalized (monochrome) image as a map from a set (called the set of points) to reals. We  define a generalized (monochrome) animation
as a map from time (discrete or continuous) to generalized images.
Linear combination of images is defined point-wise. The secondary structure on the set of points might vary. 
To obtain a conventional monochrome image the secondary structure typically assigns coordinates from
a discretized rectangle to points. For display purposes, zero is normally associated with gray level
in the middle between the most dark and the most bright possible values. This approach allows the use
of both positive and negative coefficients in the linear combinations of generalized images and animations.

Conventional color images and music are examples of generalized animations, and the use
of linear combinations of those is standard in video and audio mixing software.

One feature shared between animations and probabilistic programs is that complex behaviors
can often be expressed by very short programs. Many software experiments including 
our own work with simulated reflection in water waves~\cite{MBukatinMatthews, Fluid} demonstrate that interesting
and expressive dynamics can result from simple programs.

Another feature animations seem to share with
sampling architecture is that they tend to be non-brittle, and that their mutations and crossover
tend to produce meaningful results in the evolutionary setting~\cite{SDraves}. This architecture provides
a direct way to incorporate aesthetic criteria into software systems. This architecture
can also leverage existing animations, digital and physical (such as light refections and refractions in water), 
as computational oracles.

A lot of expressive power of this architecture comes from the ability to have non-standard secondary structures on the set of points.
Points can be associated with vertices or edges of a graph, grammar rules, positions in a matrix, etc. One should be able to formulate
mechanisms of higher-order animation programming via variable illumination of elements of such structures. 

\subsection{Negative Probability}

In order to enable both positive and negative coefficients for probabilistic sampling one can allow
sampling via two parallel sampling channels: a positive one and a negative one. 

There is evidence that signed functions are sampled via parallel positive and negative channels in
neural system, for example, in retina  (see pages 65 and 173 of~\cite{DMarr}). The idea that
some of the brain functioning might be understood as Markov Chain Monte Carlo sampling was
developed in recent years and led to fruitful applications to the computational schemes
robust with respect to noise and benefiting from presence of noise and thus suitable
for implementation in low-powered circuits (see~\cite{WMaass} and references therein).
The combination of this idea and of the evidence for sampling via parallel positive and
negative channels is suggestive.

One way to understand and formalize this situation is via allowing negative values
for probabilities and probability densities. Quasiprobability distributions allowing both
positive and negative probability values have long history. Their first prominent use
comes in phase space formulation of quantum mechanics via Wigner quasiprobability distribution
in 1940s~\cite{JMoyal,HGroenewold}. The intuition behind the notion of negative
probability is discussed in detail in~\cite{RFeynman}. More recently, negative probabilities
are finding use in quantum algorithms~\cite{CMiquelPazSaraceno}.

In denotational semantics of probabilistic programs, Dexter Kozen found it
fruitful to replace the space of probability distributions with the space of signed measures~\cite{DKozen}.
This allowed him to express denotations of probabilistic programs as 
 continuous linear operators with finite norms. The probabilistic powerdomain
was embedded into the positive cone of the resulting Banach lattice. 

\subsection{Partial Inconsistency, Non-monotonic Inference, and Vector Semantics}

Addition of the elements expressing partial degrees of contradiction results in
an embedding of an approximation domain into a vector space in yet another important case,
the interval numbers, by extending them with {\em pseudosegments} $[a,b]$ with the contradictory
property that $b < a$.

The resulting spaces tend to be equipped with two Scott topologies dual to each other,
which enables both upwards and downwards computable inference steps, and thus
facilitates non-monotonic reasoning.

The resulting mathematical landscape is a field directly adjacent to the main topic
of this paper. We review this field and present some of our own results there in Section~\ref{partial_iinconsitency}.

\subsection{Almost Continuous Transformations of Dataflow Programs}\label{almost_continuous}

Because probabilistic sampling and generalized animation are both stream-based,
dataflow programming is a natural framework for this situation. Dataflow architecture
is convenient for program learning, because syntax of dataflow programs would typically
be more closely related to their semantics than the syntax of more conventional
programs.

The ability to take linear combinations of execution runs allows us to introduce the
notion of {\em almost continuous transformation} of dataflow programs~\cite{MBukatinMatthews}.
This architecture is applicable to probabilistic sampling and to generalized animation.
We implemented an open source software prototype demonstrating the use
of those techniques for ordinary animations~\cite{Fluid}.

This architecture allows us to evolve dataflow programs in almost continuous fashion while those evolving programs
are running. This makes it possible to sample almost continuous trajectories in the space of
dataflow programs, in addition to the usual practices of sampling the syntax trees of programs, thus
enabling new evolutionary and probabilistic schemas of program learning.

\subsection{Dataflow Graphs as Matrices}\label{graphs_as_matrices}

Adopting a discipline of bipartite graphs linking nodes obtained via general transformations and nodes obtained
via linear transformations makes it possible to develop and evolve dataflow programs over these classes
of computations by {\em continuous program transformations}. The use of bipartite graphs allows us to
represent the dataflow programs from this class as matrices of real numbers and evolve and
modify programs by continuous change of these numbers~\cite{MBukatinMatthewsMatrices}.

The representation of programs as matrices of real numbers makes the task of program
learning more similar to the task of machine learning for more narrow and conventional
classes of models.

\section{Parallels between Methods of Evolutionary Programming and Probabilistic Sampling}\label{evolution_vs_sampling}

The connections between probabilistic programming and genetic programming are much tighter than it is
usually acknowledged.

Many variants of MCMC are evolutionary in spirit. Acceptance/rejection of the samples corresponds to selection.
Production of new samples via modifications of the accepted ones corresponds to mutations to produce offspring from the
survivors.

The Bayesian Optimization Algorithm changes the procedure of
producing the next generation in genetic algorithms from pairwise
crossover to resampling from the estimated distribution of the
individuums selected for fitness~\cite{MPelikan}. This scheme
of crossover is used by the seminal MOSES system~\cite{MLooks}. This is similar
in spirit to population-based methods of sampling.

\section{Some Recent Advances in Higher-Order Probabilistic Programming}\label{prob_prog}

We are seeing very rapid progress in probabilistic programming in recent years. 

What particularly catches our attention is a series of results solving various
computer vision problems as Bayesian inverse problems to computer graphics rendering, 
starting with~\cite{VMansinghkaKulkarnPeroTenenbaum}.

For an example of a powerful model learning scheme for probabilistic programs using matrix decomposition and a context-free grammar
of models see~\cite{RGrosseSalakhutdinovFreemanTenenbaum}.

The term ``higher-order probabilistic programming" usually means a higher-order functional
programming language implementing sampling semantics. Recently we are seeing examples of research
implementing higher-order sampling schemas in a more narrow and focused sense of the word:
samplers which generate other samplers, probabilistic programs sampling the space of
probabilistic programs.This is a particularly important development for program learning. 

A recent work on learning probabilistic programs within the ``sampling the samplers" paradigm
 by Perov and Wood allows, in particular, ``compilation" of probabilistic programs 
so that the resulting samplers just sample the posterior directly
without sampling the whole joint distribution (another possible name for this procedure which comes to mind
is ``partial evaluation", although neither term is quite adequate for this novel procedure).
The work is done using the new Anglican engine which implements a probabilistic programming language
similar to Venture, but is written in Clojure (which should enable better parallelization and better performance
scaling with more hardware) and uses a higher-order PMCMC (``Particle Markov Chain Monte Carlo")
sampling scheme, where efficient high-dimensional proposal distributions for MCMC are generated
by particle filters. The work follows earlier successes of Maddison and Tarlow in capturing 
frequent context-dependent syntactic patterns of code from open source repositories
within generative models (see~\cite{YPerovWood} and references therein).

Another important work done with the use of multilevel ``generative models emitting generative models"
architecture is the research by Lake in compositional concept learning~\cite{BLake}. The typical
tasks performed are learning the letters of synthetic alphabet and spoken Japanese-like words.
The author claims that this is the first time when a machine learning system combines
learning from one or a few examples (rather than from big data corpora) with 
learning rich conceptual representations.

\section{Partial Inconsistency, Non-monotonic Inference, and Vector Semantics}\label{partial_iinconsitency}

The traditional mathematical view is that there is only one kind of contradiction and that all
contradictions imply each other and everything else.
However, there is also rich tradition of studying various kinds of graded or partial contradictions.

There are a number of common motives appearing multiple times in various studies of graded inconsistency.
These common motives  link a variety of independently done studies together and serve as focal elements 
of what we call the {\em partial inconsistency landscape}~\cite{MBukatinKoppermanMatthews}. We list many of these common motives and
some of their interplay.

An especially important motive is that in the presence of partial inconsistency many otherwise impoverished
algebraic structures become groups and vector spaces. In particular, domains for denotational semantics
tend to acquire group and vector space structure when partial inconsistency is present.

Known applications include handling of inconsistent information and non-monotonic and\linebreak anti-monotonic
inference. Perhaps even more importantly for the advanced AI, vector semantics is likely to offer new powerful schemes
for program learning, as we are arguing in this paper.

We provide a necessarily incomplete overview of this field here and present some of our results. For more details,
see~\cite{MBukatinKoppermanMatthews2} and other materials in~\cite{MBukatinPIVS}.

\subsection{Focal Elements of the Partial Inconsistency Landscape}

\begin{itemize}
\item{Various forms of {\em negative measure} (negative length and distance, negative probability and signed measures, negative membership and signed multisets)}
\item{Bilattices}
\item{Bitopology}
\item{Domains with group and vector space structures}
\item{Bicontinuous domains}
\item{The domain of arrows, $D^{Op}\times D$ or $C^{Op}\times D$}
\item{Non-monotonic and anti-monotonic inference}
\item{Modal and paraconsistent logic and possible world models}
\item{Hahn-Jordan decomposition or ``bilattice pattern":\newline $x = (x \wedge 0) + (x \vee 0)$ or $x = (x \wedge \bot) \sqcup (x \vee \bot)$}
\end{itemize}

\subsection{Partially Inconsistent Interval Numbers}

Interval numbers are segments $[a,b]$ on the real line where $a\leq b$. One can extend
interval numbers by adding {\em pseudosegments} $[a,b]$ with the contradictory
property that $b < a$. This structure was independently discovered many times and is known
under various names including Kaucher interval arithmetic, directed interval arithmetic, generalized interval arithmetic,
and modal interval arithmetic (a comprehensive repository of literature on the subject is maintained by Evgenija Popova~\cite{EPopova}).
The first mention known to us is by Warmus in 1956~\cite{MWarmus}.
Our group tends to call this structure {\em partially inconsistent
interval numbers}.

There are two partial orders on partially inconsistent interval numbers.
The {\em informational order}, $\sqsubseteq$, is defined by reverse
inclusion on interval numbers: $[a,d] \sqsubseteq [b,c]$ iff $a\leq b$ and $c\leq d$.
The same formula is used for partially inconsistent interval numbers.
The {\em material order} is component-wise: $[a,b] \leq [c,d]$ iff $a \leq c$ and $b \leq d$.

Addition on interval numbers (and partially inconsistent interval numbers) is defined component-wise:
$[a_1,b_1] + [a_2,b_2] = [a_1+a_2, b_1+b_2]$.  

The operation of {\em weak minus} is defined as
$-[a,b] = [-b,-a]$. Addition and weak minus are {\em monotonic} with respect to $\sqsubseteq$.

Consider $-[a,b] + [a,b] = [-b,-a] + [a,b] = [a-b, b-a]$. If $a < b$, then the strict inequality, 
$[a-b, b-a] \sqsubset [0,0]$, holds. So if $a < b$, $-[a,b] + [a,b]$ approximates $[0,0]$, but is
not equal to it, hence interval numbers with weak minus don't form a group.

If one allows pseudosegments, one can define the component-wise {\em true minus}: $-[a,b] = [-a,-b]$. Partially
inconsistent interval numbers with the component-wise addition and the true minus form a group (and a
2D vector space over the reals). The true minus maps precisely defined numbers, $[a, a]$, to precisely
defined numbers, $[-a, -a]$. Other than that, the true minus maps segments to pseudosegments
and maps pseudosegments to segments. The true minus is {\em anti-monotonic} with respect to $\sqsubseteq$.

\subsection{Bilattices}

A bilattice is a set equipped with two lattice structures defining two partial orders, the {\em material order}, $\leq$,
and the  {\em informational order}, $\sqsubseteq$, and Ginsberg involution\footnote{an involution is a function $f$ such that $f(f(x)) = x$ for all $x$ in the domain of $f$}
monotonic with respect to $\sqsubseteq$,
anti-monotonic with respect to $\leq$, and preserving appropriate lattice structures. Additional axioms are
often imposed.

Bilattices were introduced by Matthew Ginsberg~\cite{MLGinsberg} to provide a unified framework for a variety of inferences schemes
used in AI, such as {\em non-monotonic inference}, inference with uncertainty, etc. They are now ubiquitous
in the studies of partial and graded inconsistency.

The simplest example of a bilattice is the four-valued logic: $f < \bot < t, f < \top < t$, 
$\bot \sqsubset f \sqsubset \top, \bot \sqsubset t \sqsubset \top$. 

Partially inconsistent interval numbers form a bilattice. Sometimes one wants both orders to form
complete lattices. This can be achieved by allowing $a$ and $b$ to also take $-\infty$ and $+\infty$ as values,
or by confining $a$ and $b$ within a segment $[A,B]$, in both cases sacrificing the property of partially inconsistent
interval numbers being a group.

If we consider all partially inconsistent interval numbers without infinities or allow $a$ and $b$ to take $-\infty$ and +$\infty$ values,
 or if we confine $a$ and $b$ within segment $[-A,A]$, then Ginsberg involution is the weak
minus. If we confine $a$ and $b$ within a segment $[A,B]$, then Ginsberg involution
maps $[a,b]$ to $[A+B-b, A+B-a]$. One important case here is $[A,B]=[0,1]$.

\subsection{Bitopology and Non-monotonic Inference}\label{bitopology_nonmono}

Asymmetric topology such as Scott topology generated by a partial order $\sqsubseteq$ is often used in computer science to encode monotonic inference
and limits of monotonic inference. For example, the upper topology on the real line consists of the open
rays $(a,+\infty)$ (take $a=-\infty$ and $a=+\infty$ to represent the whole space and the  empty set).
This topology encodes the processes generating monotonically non-decreasing sequences of reals,
$x_1 \leq x_2 \leq \dots$, and their limits. 

Scott continuous functions are functions respecting this structure. More specifically, Scott continuous functions
between two spaces with Scott topologies are monotonic functions preserving appropriately defined limits.
A classic exposition of these ideas is~\cite{DScottOutline}.

For an exposition of inference in Scott domains see, for example, Chapter 5 of~\cite{MBukatinThesis}. Because
Hasse diagrams depict partially ordered sets in such a fashion that the larger elements are above the smaller elements,
we say that the standard monotonic inference in Scott domains is directed upwards (the elements become
larger in the process of inference).
  
If there are two Scott topologies on the same set with associated partial orders pointing into opposite directions, one can infer both upwards and downwards,
thus enabling non-monotonic inference.

In our example, one can also consider the lower topology on the real line consisting of the open rays
$(-\infty,b)$, and this is the second Scott topology, encoding the processes generating  monotonically non-increasing sequences of reals,
$y_1 \geq y_2 \geq \dots$, and their limits. Switching between these two topologies one can
encode non-monotonic sequences.

A space with two topologies is called a bitopological space, and a
space with Scott topologies generated by $\sqsubseteq$ and $\sqsupseteq$ with certain additional
properties is called a bicontinuous domain~\cite{KKeimel}.

\subsection{Order Reversal and the Domain of Arrows}

Consider a bicontinuous domain, $(X,\sqsubseteq, \sqsupseteq)$.  The partial order $\sqsupseteq$ which is the
order opposite to $\sqsubseteq$ then defines the dual space, $X^{Op} = X^* = (X, \sqsupseteq, \sqsubseteq)$, which is also a bicontinuous domain.

For a bicontinuous domain $(X,\sqsubseteq, \sqsupseteq)$ we say that $\sqsubseteq$ is pointing upwards
(``the main partial order of the space") and $\sqsupseteq$ is pointing downwards (``the auxiliary or dual
order of the space").

If we think informally about an arrow from space $X$ to space $Y$, then our intuition tells us that the arrow
is greater if it ``points more upwards", that is, if its right end is higher, and its left end is lower.

Formalizing this intuition we define the space of arrows from $X$ to $Y$ as $X^* \times Y$.

If we consider real numbers $\mathbb{R}$ with the standard order, $\sqsubseteq = \leq$, then partially inconsistent interval
numbers are a space of arrows pointing from the right ends of the segments to the left ends of the segments,
$\mathbb{R} \times \mathbb{R}^*$.

If $\mathbb{R}$ is modified to become a domain, $R$ (by adding $-\infty$ and $+\infty$ as values or by taking a finite segment),
we call $R \times R^*$ a domain of arrows.

There are two ways to describe Scott topology in terms of generalized distances. One is via asymmetric quasi-metrics, with $d(x,y)=0$ if and only if $x \sqsubseteq y$.
Another is via dropping the $d(x,x)=0$ requirement which leads to relaxed and partial metrics. Quasi-metrics of this kind are monotonic with respect to one of
the variables and anti-monotonic with respect of another variable. So the only way to have these generalized distances to be Scott continuous as functions
from $X \times X$ to the domain representing distances is via the route of relaxed and partial metrics~\cite{MBukatinScott}.

In the bicontinuous situation, quasi-metrics can be understood as (Scott continuous) order-preserving maps from the domain of arrows, $X^* \times X$, to the
domain representing distances.

Order-reversing involutions ($x \sqsubseteq y \Leftrightarrow f(y) \sqsubseteq f(x)$ and $f(f(x))=x$) play a prominent role in this context.
From the viewpoint of domain theory, order-reversing involutions should be thought of as order-preserving maps $X \rightarrow X^*$ (or $X^* \rightarrow X$). Hence order-reversing
involutions are order-preserving maps $X^* \times X  \rightarrow X \times X^*$ (and vice versa) on the domain of arrows. 

\subsection{Bitopology and Partial Inconsistency}

There are at least three ways bitopologies occur in
studies of partial inconsistency. The connections between partial inconsistency and bitopological Stone duality
via the notion of $d$-frame (Jung-Moshier frame) are explored in~\cite{AJungMoshier} (see also~\cite{JLawson}). A fuzzy bitopology valued in lattice $L$ is
a fuzzy topology valued in the bilattice $L^2$ (in particular, an ordinary bitopology is a topology valued in
the four-valued logic)~\cite{SRodabaugh}. Finally, in the context of bitopological groups and the anti-monotonic group inverse
the following situation is typical: two topologies, $T$ and $T^{-1}$, are group dual
of each other (i.e. the group inverse induces a bijection between the respective systems of open sets), 
the multiplication is continuous with respect to both topologies, and the group inverse
is a bicontinuous map from $(X,T,T^{-1})$ to its bitopological dual, $(X,T^{-1},T)$~\cite{SAndimaKoppermanNickolas}.

All these motives are prominent for the case of partially inconsistent interval numbers~\cite{MBukatinKoppermanMatthews2}.

\paragraph{D-frames.} Partially inconsistent interval numbers over reals extended with $\pm\infty$ are isomorphic
to the $d$-frame of the (lower, upper) bitopology on the reals.

Consider the (lower, upper) bitopology on the real line, that is the bitopology where the first
topology is the lower topology, and the second topology is
the upper topology (see Section~\ref{bitopology_nonmono}). Define the bilattice isomorphism
between the d-frame elements, i.e. pairs $\langle L,U \rangle$ of the respective open sets, and
partially inconsistent interval numbers. A pair $\langle L,U \rangle$ is a pair of open rays,  $\langle (-\infty,a),(b,+\infty) \rangle$
($a$ and $b$ are allowed to take $-\infty$ and +$\infty$ as values).
This pair corresponds to a partially inconsistent interval number $[a,b]$. Consistent,
i.e. non-overlapping, pairs of open rays ($a\leq b$) correspond to segments.
Total, i.e. covering the whole space, pairs of open rays ($b < a$) correspond to pseudosegments.

\paragraph{Group dual topologies.} The minus operation on real numbers is bicontinuous from the (lower, upper) bitoplogy to the (upper, lower) bitopology
and vice versa. The corresponding map between the d-frames is very similar to the weak minus (Ginsberg involution), except
that the order of bitopological components also needs to be swapped to respect bitopological duality in this case
(partially inconsistent interval numbers are a Cartesian product of lower and upper bounds; swapping can
be thought of as changing the order of components in this Cartesian product).

In a similar fashion, the true minus operation on the partially inconsistent interval numbers is bicontinuous between
a ($T,T^{-1}$) bitopology on the partially inconsistent interval numbers and its dual ($T^{-1},T$) bitopology. (Here $T$ and
$T^{-1}$ must be group dual topologies of each other, e.g. the Scott topology corresponding to $\sqsubseteq$ and
the Scott topology corresponding to $\sqsupseteq$.)

\paragraph{Rodabaugh correspondence.} Any real-valued fuzzy
bitopology can be represented as fuzzy topology valued in partially inconsistent interval numbers.

The open sets of the upper topology on the reals is a particular representation of
real numbers extended with $\pm\infty$, the same is true about the
open sets of the lower topology on the reals. Consider a particular
form of  real-valued fuzzy bitopology, namely the multivalued bitopology where the first multivalued
topology is valued in the lower topology, and the second multivalued topology is
valued in the upper topology, that is a (lower, upper)-valued bitopology.

Consider the following mild generalization of the correspondence described in~\cite{SRodabaugh}. 
$(L,M)$-valued bitopology can be understood as $L \times M$-valued topology via $L^X \times M^X \cong (L \times M)^X$ isomorphism.

Hence the (lower, upper)-valued bitopology can be understood as
the topology valued in the (lower, upper) d-frame,
i.e. the topology valued in  partially inconsistent interval numbers over reals extended with $\pm\infty$.
Further discussion of the intuition involved here is in the slides 37-38 of~\cite{MBukatinKoppermanMatthews2}.

\subsection{Paraconsistent Version of Fuzzy Mathematics}

It seems that mathematics of partial inconsistency should be bilattice-valued. The Rodabaugh
correspondence is one of the indications of that, as $L \times M$ is naturally a bilattice, 
with the informational order, $\sqsubseteq_{L \times M}$, being obtained from the product $(L, \sqsubseteq_L) \times (M, \sqsubseteq_M)$
and the material order, $\leq_{L \times M}$, being obtained from the product of $L$ by the dual of $M$, $(L, \sqsubseteq_L) \times (M, \sqsupseteq_M)$.

While the fuzzy mathematics in general is lattice-valued, the situations where the lattice is $[0,1]$ or otherwise based on real numbers
remain important. Similarly, while mathematics of partial inconsistency is in general likely to be valued in bilattices, the particular situations where the bilattice
is based on partially inconsistent real numbers (whether confined within $[0,1]$, $[-1,1]$, or $[-\infty, +\infty]$) are likely to play
important roles. 

The paraconsistent equivalent of real-valued fuzzy mathematics is mathematics valued
in partially inconsistent interval numbers.

\subsection{Partial and Relaxed Metrics}

The standard partial metric on the interval numbers is $p([a_1,b_1],[a_2,b_2]) = \max(b_1,b_2) - \min(a_1,a_2)$~\cite{MBukatinKoppermanMatthewsPajoohesh}.
Hence the self-distance for $[a,b]$ is $b-a$. If we extend this formula to pseudosegments, the self-distance of pseudosegments turns out to be negative.

Partial metrics can be understood as upper bounds for ``ideal distances". One often has to trade the tightness of those bounds
for nicer sets of axioms. E.g. the natural upper bound for the distance between $[0,2]$ and $[1,1]$ is 1, and there is a weak partial
metric which yields that. However, if one wants to enjoy the axiom of small self-distances, $p(x,x) \leq p(x,y)$, one has to accept
$p([0,2],[1,1]) = 2$, since $p([0,2],[0,2]) = 2$.

A similar trade can be made for lower bounds. The standard interval-valued relaxed metric produces the gap between non-overlapping
segments as their lower bound, but takes 0 as the lower bound for the distance between overlapping segments (hence 0 is also
the lower bound for self-distance). If one settles for a less tight lower bound and allows the lower bound to be negative in those cases, one can obtain a distance with
much nicer properties: $l([a_1,b_1],[a_2,b_2]) = \max(a_1,a_2) - \min(b_1,b_2)$.

We think about the pair $\langle l, p \rangle$ as a relaxed metric valued in partially inconsistent interval numbers. The self-distance of
$[a,b]$ is $[a-b, b-a]$ and the self-distance of a pseudosegment is a pseudosegment. 

The map $[a,b] \mapsto [b,a]$ expressing the symmetry between segments and pseudosegments also transforms $\langle l, p \rangle$
into $\langle p, l \rangle$.

\subsection{Signed Measures and Signed Multisets}

One way to think about $p([a,b],[a,b]) = b-a$ is to say that a pseudosegment has a negative length.

We can also revisit the correspondence between the elements of the (lower, upper) bitopology d-frame,
$\{\langle (-\infty,a),(b,+\infty) \rangle\}$, and the partially inconsistent interval numbers. Consider the characteristic
function mapping the real line to 1 and subtract from it the characteristic functions of $(-\infty,a)$ and $(b,+\infty)$.
If $[a,b]$ is a segment, the result is the characteristic function of that segment (valued 1 for the points belonging
to the segment and 0 for the points outside the segment). If $[a,b]$ is a pseudosegment and if we allow for the overlap
between $(-\infty,a)$ and $(b,+\infty)$ to be subtracted twice, the result is the generalized characteristic function, which
equals to -1 in the open interval $(b,a)$ and equals to 0 outside $(b,a)$. So we obtain a signed multiset here allowing negative
degree of membership.

This construction is topologically
asymmetric in the following sense.
Algebraically we can say that totally defined numbers $[a, a]$ belong
to both segments and pseudosegments, or to neither.
But topologically (and via characteristic functions), this symmetry
must be broken.
We break it in favor of the ``natural" viewpoint: totally defined
numbers are segments, and not pseudosegments.
But one could also break it in favor of the dual viewpoint, by considering
dual d-frames of closed sets (and stipulating that characteristic
functions of segments take value 1 only on their interiors).

\subsection{Negative Probability and Vector Semantics}

One can think about probabilistic programs as transformers from the probability distributions on
the space of inputs to the probability distributions on the space of outputs. Dexter Kozen showed that it is
fruitful to replace the space of probability distributions by the space of signed measures~\cite{DKozen}. One
defines $\nu < \mu$ iff $\mu - \nu$ is a positive measure. The space
of signed measures is a vector lattice (a Riesz space) and a Banach space, so people call this structure a Banach
lattice. Denotations of programs are continuous linear operators with finite norms. The probabilistic powerdomain
is embedded into the positive cone of this Banach lattice. The structure of Hilbert space on signed measures can be obtained via
reproducing kernel methods (see Chapter 4 of~\cite{ABerlinetThomasAgnan}).

\subsection{Hahn-Jordan Decomposition and the Bilattice Pattern:\newline $x = (x \wedge 0) + (x \vee 0)$ or $x = (x \wedge \bot) \sqcup (x \vee \bot)$}

The Hahn-Jordan decomposition, $\mu^+ = \mu \vee 0, \mu^- = \mu \wedge 0, \mu = \mu^+ + \mu^-$, holds, due to the fact that
$x = (x \wedge 0) + (x \vee 0)$ is a theorem for all lattice-ordered groups. 

Defining $\nu \sqsubseteq \mu$ iff  $\nu^+ \leq \mu^+$ and  $\nu^- \leq \mu^-$,
one also obtains $ \mu = \mu^+ \sqcup \mu^-$, making this an instance of the ``bilattice pattern", 
$x = (x \wedge \bot) \sqcup (x \vee \bot)$. The ``bilattice pattern" appears independently in a variety
of studies on partial inconsistency~\cite{MBukatinKoppermanMatthews}.

It looks like the right degree of generality here might be lattice-ordered monoids with an extra axiom,
$x = (x \wedge 0) + (x \vee 0)$.

\subsection{Possible Worlds Indexed by Measures}

William Wadge explored various ways to  index possible worlds in the context of  intensional logic and data flow programming~\cite{WWadge}

In our case, possible worlds would be indexed by measures, which is quite attractive and feels natural: a world is distinguished by
 how often one observes various phenomena, and we do sampling observations to figure out what kind of world we currently inhabit.

In a classical situation one would normally consider probability measures for this role, but in a quantum situation
signed quasi-probability distributions or complex-valued amplitudes would naturally play this role.

\subsection{Distances Between Programs}

On one hand, Anthony Seda and M\'{a}ire Lane note that there is a natural norm for Kozen semantic spaces, which allows us to define
a conventional metric on program denotations and hence a conventional distance between programs~\cite{ASedaLane}.

On the other hand, in the context where everything is a function of a measure (possible worlds are indexed by measures),
the standard constructions of generalized distances (partial metrics, relaxed metrics, and quasi-metrics) over Scott domains which tend to be parametrized by 
measures (see, for example, Section 7 of~\cite{MBukatinKoppermanMatthewsFuzzy}) 
look quite natural. 

We tend to view the dependency of those constructions on a measure as an obstacle which needs to be overcome, but perhaps it is actually a desirable feature.

\subsection{Computational Models with Involutions}

We are currently looking at various computational models involving involutions.

Given a domain of arrows $X \times X^*$, a sequence
$(x_1, y_1), (x_2, y_2), \dots $ is called a monotonic sequence with involutive steps, 
if for any $n \in \mathbb{N}$ either $x_n \sqsubseteq x_{n+1}$ and $y_n \sqsupseteq y_{n+1}$ 
(in which case the step $n$ is called monotonic) 
or $x_n = y_{n+1}$ and $y_n = x_{n+1}$ (in which case the step $n$ is called an involution).

One can define a notion of convergence robust with respect to the insertion of pairs
of involutive steps and prove that if $(x,y)$ is a limit under this notion, then $x=y$.

Architectures based on involutive steps are rather prominent in the context of
reversible and quantum computations. For example, the well-known Grover's
quantum algorithm can be described as a sequence of reflections of subsets of a plane~\cite{ElRieffelPolak}. 

Architectures where the state of an abstract machine is an image on the plane,
and an involutive computational step selects a line on this plane and a subset symmetric
with respect to this line, and performs a reflection of the image within this subset,
seem to be quite attractive in the context of classical computations as well.

\section{Conclusion}

It is possible to talk about linear combinations of probabilistic programs when their semantics is expressed as linear operators~\cite{DKozen}.

For $0 < \alpha < 1$ and {\bf random} being a generator of uniformly distributed reals between 0 and 1, the linear operator corresponding to the program
 {\bf if random} $ < \alpha$ {\bf then P else Q} is
a linear combination of the linear operators corresponding to programs {\bf P} and {\bf Q} with coefficients $\alpha$ and $1-\alpha$. 

However, when one aims for better schemes of program learning, the situations where one can consider linear combinations of single execution runs
rather than linear combinations of the overall program meanings should be especially attractive.
In this paper we consider two such architectures, probabilistic sampling and generalized animation, 
and the recent progress in this field looks very promising.

We give an overview of mathematical material tightly connected to linear models of computations
via the partial inconsistency landscape. Vector semantics is an integral part of the
partial inconsistency landscape, and we expect that other key elements of
that landscape will be finding more uses 
as linear models of computations and their applications are further explored.

We would like to conclude by describing a possible hybrid approach to program learning.
Instead of implementing everything in terms of architectures admitting linear combinations of single execution runs one
can use a hybrid approach, mixing these architectures and traditional
software. In this context we might be inspired by hybrid hardware connecting live neural tissue and
electronic circuits.

One might decide to use large existing software components and try
to automate the process of connecting them together using 
flexible probabilistic connectors. Here one should note the progress
in automated generation of test suites for software systems.

Another hybrid approach involves the use of small inflexible components inside the 
flexible ``tissue" of linear models. Our experiments in dataflow programming with
streams supporting the notion of a linear combination of streams 
described in Sections~\ref{almost_continuous}, \ref{graphs_as_matrices} are
examples of this hybrid approach~\cite{MBukatinMatthews, MBukatinMatthewsMatrices}.
In particular, the template operations play the role of small inflexible components
in~\cite{MBukatinMatthewsMatrices}, where dataflow graphs are represented
by matrices of real numbers describing the flexible connectivity patterns from the outputs to the
inputs of a potentially countable number of template operations.

{\bf Acknowledgments.} We would like to thank Ralph Kopperman, Lena Nekludova, Leon Peshkin, Josh Tenenbaum, and Levy Ulanovsky for helpful discussions.

\label{sect:bib}
\bibliographystyle{plain}
\bibliography{LinearModels}

\begin{thebibliography}{10}

\bibitem{SAndimaKoppermanNickolas}
S.~Andima, R.~Kopperman, and P.~Nickolas.
\newblock An asymmetric {Ellis} theorem.
\newblock {\em Topology and Its Applications}, 155:146--160, 2007.

\bibitem{ABerlinetThomasAgnan}
A.~Berlinet and C.~Thomas-Agnan.
\newblock {\em Reproducing Kernel Hilbert Spaces in Probability and
  Statistics}.
\newblock Kluwer Academic Publishers, Boston, 2004.

\bibitem{MBukatinPIVS}
M.~Bukatin.
\newblock Partial inconsistency and vector semantics of programming languages
  (a repository of research materials).
\newblock \url{http://www.cs.brandeis.edu/~bukatin/partial_inconsistency.html}.

\bibitem{MBukatinThesis}
M.~Bukatin.
\newblock {\em Mathematics of Domains}.
\newblock PhD thesis, Brandeis University, 2002.
\newblock \url{http://www.cs.brandeis.edu/~bukatin/Mathematics-of-Domains.pdf}.

\bibitem{MBukatinKoppermanMatthews}
M.~Bukatin, R.~Kopperman, and S.~Matthews.
\newblock Partial inconsistency landscape: an overview. {28th Summer Conference
  on Topology and Its Applications, Nipissing University, July, 2013}.
\newblock
  \url{http://www.cs.brandeis.edu/~bukatin/PartialInconsistencyJul24.pdf}.

\bibitem{MBukatinKoppermanMatthews2}
M.~Bukatin, R.~Kopperman, and S.~Matthews.
\newblock Progress report on partial inconsistency, bitopology, and vector
  semantics. {Conference on Computational Topology and Its Applications, Kent
  State University, Ohio, November 2014}.
\newblock
  \url{http://www.cs.brandeis.edu/~bukatin/PartialInconsistencyProgressNov2014.pdf}.

\bibitem{MBukatinKoppermanMatthewsFuzzy}
M.~Bukatin, R.~Kopperman, and S.~Matthews.
\newblock Some corollaries of the correspondence between partial metrics and
  multivalued equalities.
\newblock {\em Fuzzy Sets and Systems}, 256:57--72, 2014.

\bibitem{MBukatinKoppermanMatthewsPajoohesh}
M.~Bukatin, R.~Kopperman, S.~Matthews, and H.~Pajoohesh.
\newblock Partial metric spaces.
\newblock {\em American Mathematical Monthly}, 116:708--718, 2009.

\bibitem{MBukatinMatthews}
M.~Bukatin and S.~Matthews.
\newblock Almost continuous transformations of software and higher-order
  dataflow programming, Preprint (2015).
\newblock \url{http://www.cs.brandeis.edu/~bukatin/HigherOrderDataFlow.pdf}.

\bibitem{MBukatinMatthewsMatrices}
M.~Bukatin and S.~Matthews.
\newblock Dataflow graphs as matrices and programming with higher-order matrix
  elements, Preprint (2015).
\newblock
  \url{http://www.cs.brandeis.edu/~bukatin/DataFlowGraphsAsMatrices.pdf}.

\bibitem{MBukatinScott}
M.~Bukatin and J.~Scott.
\newblock Towards computing distances between programs via {Scott} domains.
\newblock In S.~Adian and A.~Nerode, editors, {\em Logical Foundations of
  Computer Science}, LNCS, vol. 1234, pages 33--43. Springer, 1997.

\bibitem{SCarrol}
S.~Carrol.
\newblock Evo-devo and an expanding evolutionary synthesis: A genetic theory of
  morphological evolution.
\newblock {\em Cell}, 134:25--36, 2008.

\bibitem{SDraves}
S.~Draves.
\newblock The {Electric Sheep} screen-saver: A case study in aesthetic
  evolution.
\newblock In F.~Rothlauf et~al., editors, {\em Applications of Evolutionary
  Computing}, LNCS, vol. 3449, pages 458--467. Springer, 2005.
\newblock \url{http://draves.org/evomusart05/}.

\bibitem{RFeynman}
R.~Feynman.
\newblock Negative probability.
\newblock In F.~Peat and B.~Hiley, editors, {\em Quantum Implications : Essays
  in Honour of David Bohm}, pages 235--248. Routledge and Kegan Paul, 1987.
\newblock \url{http://cds.cern.ch/record/154856/files/pre-27827.pdf}.

\bibitem{Fluid}
Fluid.
\newblock Project ``{Fluid}" github repository, 2015.
\newblock \url{https://github.com/anhinga/fluid}.

\bibitem{JGerhartKirschner}
J.~Gerhart and M.~Kirschner.
\newblock The theory of facilitated variation.
\newblock {\em Proc. Natl. Acad. Sci.}, 104(Suppl 1):8582--8589, 2007.

\bibitem{MLGinsberg}
M.~Ginsberg.
\newblock Multivalued logics: a uniform approach to inference in artifcial
  intelligence.
\newblock {\em Computational Intelligence}, 4(3):256--316, 1992.

\bibitem{HGroenewold}
H.~Groenewold.
\newblock On the principles of elementary quantum mechanics.
\newblock {\em Physica}, 12:405--460, 1946.

\bibitem{RGrosseSalakhutdinovFreemanTenenbaum}
R.~Grosse, R.~Salakhutdinov, W.~Freeman, and J.~Tenenbaum.
\newblock Exploiting compositionality to explore a large space of model
  structures, 2012.
\newblock \url{http://arxiv.org/abs/1210.4856}.

\bibitem{AJungMoshier}
A.~Jung and M.A. Moshier.
\newblock On the bitopological nature of {Stone} duality.
\newblock Technical Report CSR-06-13, School of Computer Science, University of
  Birmingham, 2006.
\newblock
  \url{http://www.cs.bham.ac.uk/~axj/pub/papers/Jung-Moshier-2006-On-the-bitopological-nature-of-Stone-duality.pdf}.

\bibitem{KKeimel}
K.~Keimel.
\newblock Bicontinuous domains and some old problems in domain theory.
\newblock {\em Electronic Notes in Theoretical Computer Science}, 257:35--54,
  2009.

\bibitem{MKirschnerGerhart}
M.~Kirschner and J.~Gerhart.
\newblock {\em The Plausibility of Life: Resolving Darwin's Dilemma}.
\newblock Yale University Press, 2005.

\bibitem{DKozen}
D.~Kozen.
\newblock Semantics of probabilistic programs.
\newblock {\em Journal of Computer and System Sciences}, 22(3):328--350, 1981.

\bibitem{BLake}
B.~Lake.
\newblock {\em Towards More Human-like Concept Learning in Machines:
  Compositionality, Causality, and Learning-to-learn}.
\newblock PhD thesis, MIT, 2014.
\newblock \url{http://cims.nyu.edu/~brenden/LakePhDThesis.pdf}.

\bibitem{JLawson}
J.~Lawson.
\newblock Stably compact spaces.
\newblock {\em Mathematical Structures in Computer Science}, 21(1):125--169,
  2011.

\bibitem{MLooks}
M.~Looks.
\newblock {\em Competent Program Evolution}.
\newblock PhD thesis, Washington University in St. Louis, 2002.
\newblock \url{http://metacog.org/doc.html}.

\bibitem{WMaass}
W.~Maass.
\newblock Noise as a resource for computation and learning in networks of
  spiking neurons.
\newblock {\em Proc. of the IEEE}, 102(5):860--880, 2014.

\bibitem{VMansinghkaKulkarnPeroTenenbaum}
V.~Mansinghka, T.~Kulkarni, Yu. Perov, and J.~Tenenbaum.
\newblock Approximate {Bayesian} image interpretation using generative
  probabilistic graphics programs, 2013.
\newblock \url{http://probcomp.csail.mit.edu/gpgp}.

\bibitem{DMarr}
D.~Marr.
\newblock {\em Vision}.
\newblock W. H. Freeman and Company, New York, 1982.

\bibitem{CMiquelPazSaraceno}
C.~Miquel, J.~Paz, and M.~Saraceno.
\newblock Quantum computers in phase space.
\newblock {\em Phys. Rev. A}, 65(6):062309, 2002.
\newblock \url{http://arxiv.org/abs/quant-ph/0204149}.

\bibitem{JMoyal}
J.~Moyal.
\newblock Quantum mechanics as a statistical theory.
\newblock {\em Proceedings of the Cambridge Philosophical Society}, 45:99--124,
  1949.
\newblock \url{http://physics.bu.edu/~youssef/quantum/moyal.pdf}.

\bibitem{MPelikan}
M.~Pelikan.
\newblock {\em Bayesian Optimization Algorithm: from Single Level to
  Hierarchy}.
\newblock PhD thesis, University of Illinois at Urbana-Champaign, 2002.
\newblock \url{http://www.medal-lab.org/files/2002023.pdf}.

\bibitem{YPerovWood}
Yu. Perov and F.~Wood.
\newblock Learning probabilistic programs, 2014.
\newblock \url{http://arxiv.org/abs/1407.2646}.

\bibitem{EPopova}
E.~Popova.
\newblock The arithmetic on proper \& improper intervals (a repository of
  literature on interval algebraic extensions).
\newblock \url{http://www.math.bas.bg/~epopova/directed.html}.

\bibitem{ElRieffelPolak}
E.~Rieffel and W.~Polak.
\newblock {\em Quantum Computing: A Gentle Introduction}.
\newblock MIT Press, 2011.

\bibitem{SRodabaugh}
S.~Rodabaugh.
\newblock Functorial comparisons of bitopology with topology and the case for
  redundancy of bitopology in lattice-valued mathematics.
\newblock {\em Applied General Topology}, 9(1):77--108, 2008.

\bibitem{DScottOutline}
D.~Scott.
\newblock Outline of a mathematical theory of computation.
\newblock Technical Report PRG02, Oxford University Computing Laboratory, 1970.
\newblock \url{https://www.cs.ox.ac.uk/files/3222/PRG02.pdf}.

\bibitem{ASedaLane}
A.~Seda and M.~Lane.
\newblock On continuous models of computation: towards computing the distance
  between (logic) programs.
\newblock In J.~Morris, B.~Aziz, and F.~Oehl, editors, {\em 6th International
  Workshop on Formal Methods (IWFM'03 Proceedings)}. Dublin City University,
  2003.
\newblock \url{http://www.bcs.org/upload/pdf/ewic_iwfm03_paper1.pdf}.

\bibitem{WWadge}
W.~Wadge.
\newblock Intensional logic in context.
\newblock In M.~Gergatsoulis and P.~Rondogiannis, editors, {\em Intensional
  Programming II: based on the papers at Islip 99}, pages 1--13. World
  Scientific, 2000.
\newblock
  \url{http://www.cse.unsw.edu.au/~plaice/archive/ISLIP/ip2/ip2_001.pdf}.

\bibitem{MWarmus}
M.~Warmus.
\newblock Calculus of approximations.
\newblock {\em Bull. Acad. Pol. Sci., Cl. III}, 4(5):253--259, 1956.
\newblock \url{http://www.cs.utep.edu/interval-comp/warmus.pdf}.

\end{thebibliography}

\end{document}